\documentclass[final,5p,times,twocolumn]{elsarticle}

\usepackage{amssymb}

\usepackage[latin1]{inputenc}
\usepackage{amsmath}
\usepackage{amsfonts}
\usepackage{amssymb}
\usepackage{graphicx}
\usepackage{subfig}
\usepackage{mathrsfs}
\usepackage{listings}
\usepackage{verbatim}
\usepackage{mathtools}
\usepackage{simplewick}
\usepackage{color}
\usepackage{subfig}

\newcommand{\be}{\begin{equation}}
\newcommand{\ee}{\end{equation}}
\newcommand{\bq}{\begin{eqnarray}}
\newcommand{\eq}{\end{eqnarray}}

\newcommand{\one}{\hbox{\rm 1\kern-.27em I}}

\journal{Physics Letters B}

\begin{document}

\begin{frontmatter}


\author{Paul Mansfield \\Centre for Particle
Theory, Department of Mathematical Sciences, University of
Durham, Durham DH1 3LE, UK \\ Email:
P.R.W.Mansfield@durham.ac.uk}

\title{The Fermion Content of the Standard Model   
        \\DCPT-14/51
        }


\author{}

\address{}

\begin{abstract}
%The world-line approach to chiral fermions is generalised to construct a simple model that automatically generates the sum over gauge group representations and chiralities of a single generation of fermions in the Standard Model, augmented by a sterile neutrino. 

We describe a simple model that automatically generates the sum over gauge group representations and chiralities of a single generation of fermions in the Standard Model, augmented by a sterile neutrino. The model is a modification of the world-line approach to chiral fermions.

\end{abstract}


\end{frontmatter}


\section{Introduction}
The natural language of high energy physics is second quantisation, or quantum field theory, providing a simple formulation of the processes of particle creation and annihilation. Its triumphs are the astonishing accuracy of QED and the succinctness and elegance of the Standard Model. However, first quantisation can also offer significant insights, which is perhaps not surprising given the success of the approach in studying string theory. For example, Strassler \cite{Strassler:1992zr} showed that the world-line formalism could be used to derive the Bern-Kosower rules that streamline perturbative calculations. For a more recent application of the formalism see \cite{Ilderton:2014mla}. In this paper we will find that a simple modification offers the possibility of unifying quarks and leptons into a single mathematical structure. In particular we will show that the sum over gauge group representations and chiralities of a generation of the Standard Model arises automatically. We begin by describing 
a world-line approach to chiral fermions and then set out our model.

\section{World-line description of chiral fermions}
The action for a left-handed fermion moving in a background gauge-field $A$ is $S=\int d^4x\,i\xi^\dagger \bar\sigma\cdot D\xi$ in the representation \cite{Herbie} of the Dirac matrices 
\be (\gamma^\mu)=\left( \begin{array}{cc}
0 & \sigma^\mu  \\
\bar\sigma^\mu & 0  \end{array} \right),\quad
(\sigma^\mu)=(1_2,\sigma^i),\quad (\bar\sigma^\mu)=(1_2,-\sigma^i)\,.
\ee
$D=\partial+A$ and the coupling has been absorbed into the gauge-field. The effective action obtained by integrating out the fermions is a functional of the backgound. Rather than study this directly it has often been found more convenient to consider its variation under a change of $A$, as in  \cite{AlvarezGaume:1985xf}-\cite{Mondragon:1995ab}:
\be
\delta \log \int {\mathscr{D}}(\bar\xi,\xi)\, e^{iS}=
{\rm Tr} \,\left((\bar\sigma\cdot D)^{-1}\bar\sigma\cdot \delta A\right)
\ee
which can be written in terms of the full Dirac matrices as:
\be
{\rm Tr} \,\left((\sigma\cdot D\,\bar\sigma\cdot  D)^{-1}\sigma\cdot D\,\bar\sigma\cdot \delta A \right)\nonumber
\ee
\be
=-\int_0^\infty dT\,{\rm Tr} \,\left(P_L\,\exp \left(T(\gamma\cdot D)^2\right)\gamma\cdot D\,\gamma\cdot  \delta A\right)\label{spin}
\ee
%\bq{\rm Tr} \,\left((\sigma\cdot D\,\bar\sigma\cdot  D)^{-1}\sigma\cdot \delta A\,\bar\sigma\cdot  D\right)&=&{\rm Tr} \,\left(P_L\,\left((\gamma\cdot D)^2\right)^{-1}\gamma\cdot \delta A\,\gamma\cdot  D\right)\nonumber\\&=&\int_0^\infty dT\,{\rm Tr} \,\left(P_L\,\exp \left(T(\gamma\cdot D)^2\right)\gamma\cdot \delta A\,\gamma\cdot  D\right)\eq
with $P_{L}=(1-\gamma_5)/2$. 
$\gamma$-matrices can be represented by functional integrals over anti-commuting variables $\psi$:
\be
\int{\mathscr{D}}\psi 
%\,{\cal N}
\,e^{-\int_0^{2\pi} dt \left(\frac{1}{2}\psi\cdot \dot\psi-{\sqrt 2}\eta\cdot\psi\right)}\propto {\cal P}\,{\rm tr}\,e^{\,\int_0^{2\pi} dt \,\eta\cdot\gamma}
\ee
where $\eta$ is an anti-commuting source, and we impose anti-periodic boundary conditions $\psi(2\pi)=-\psi(0)$. We write $\propto$ as the two sides differ by a (possibly divergent) normalisation constant. Periodic boundary conditions produce a $\gamma_5$ insertion resulting in ${\cal P}\,{\rm tr}\,\gamma_5 e^{\,\int_0^{2\pi} dt\, \eta\cdot\gamma}$. Now
$(\gamma\cdot D)^2=D^2+F_{\mu\nu}\gamma^\mu\gamma^\nu/2$.
Using this, together with the Wick-rotated path integral representation of the heat kernel\cite{Schubert:2001he} $\exp \left(T(\gamma\cdot D)^2\right)$ leads to the world-line representation of (\ref{spin}) which sums over closed paths $w^\mu(t)$:
\bq
-\int_0^\infty {dT}\int_{\rm ap-p}
{\mathscr{D}}(w,\psi)\,{\cal N}\,e^{-\frac{1}{2}\int_0^{2\pi} dt \left(\frac{\dot w^2}{T}+\psi\cdot \dot\psi\right)}
\nonumber\\
\times\, {\cal P}\,{\rm tr}\,\left(g(2\pi)\,\int_0^{2\pi} dt\,\psi\cdot\dot w
\,\psi\cdot \delta A/T\right)
\label{2}
\eq
where 
\be
g(t')={\cal P}\,{\rm exp}\left(-\int_0^{t'} {\cal A}^RT_R\,dt\right)\,,\quad {\cal A}={\dot w} \cdot A+\frac{T}{2}F_{\mu\nu}\psi^\mu\psi^\nu \,,
\label{3}
\ee
and $\{T_R\}$ are anti-Hermitian Lie algebra generators. ${\rm tr} \,g(2\pi)$ is the (super)-Wilson loop. ${\cal N}$ is a normalisation constant. The action is a gauge-fixed version of \cite{Brink:1976uf}. The ${\rm ap-p}$ subscript on $\int$ denotes that the integral with periodic boundary conditions on the $\psi$ should be subtracted (with appropriate choices of ${\cal N}$) from that with anti-periodic conditions so as to generate the $P_L$ insertion.
Summing the two contributions instead generates the expression for a right-handed fermion. This representation can be interpreted as the contribution of a particle with world-line $w^\mu(t)$ traced out in the direction of increasing $t$, but, corresponding to re-writing the Lagrangian in the form $(\xi ^t\sigma_2)i\sigma\cdot (\partial-A^t)(\sigma_2\xi^*)$, it can equally be interpreted as the contribution of an anti-particle moving around the closed path in the opposite sense because the effect of changing direction on $g(2\pi)$ is to change the sign of $A$ and replace path-ordering by anti-path-ordering, i.e. $g\rightarrow g^\dagger$, and interchange chiralities.

To compute the full one-loop effective action for the fermions in the Standard Model we would have to sum (\ref{2}) over the appropriate representations of SU(2) and SU(3) and weak hypercharges and take correlated combinations of periodic and anti-periodic boundary conditions on the $\psi$ to pick out the corresponding chiralities. For each generation this consists of ten pieces: the leptons form a left-handed $SU(2)$ doublet, $E_L=(\nu,l)^t$ (in the notation of \cite{Bailin}) and a right-handed $SU(2)$ singlet, $l_R$, both of which are $SU(3)$ singlets. The quarks form a left-handed $SU(2)$ doublet, $Q_L=(U,D)^t$, and two right-handed singlets $U_R$ and $D_R$, all being in the fundamental representation of $SU(3)$. To these can be added the anti-particles, transforming in conjugate representations of $SU(3)$ and with opposite chiralities. The $U(1)$ charges associated to these species are $-1/2$, $-1$, $1/6$, $2/3$ and $-1/3$ respectively with the anti-particles taking the opposite $U(1)$ charges. We denote 
anti-particles by bars: $\bar E_L$, $\bar l_R$, etc. The point of 
this paper is to demonstrate that this somewhat complicated sum can be written very simply as an integral over additional fermionic variables on the world-line.

 \section{The model}
It is well known that if $\tilde\phi_r$ and $\phi_s$ are a set of anti-commuting operators with $\{\tilde\phi_r,\,\phi_s\}=\delta_{rs}$ then the operators $\hat T^R\equiv\tilde\phi_r\, T^R_{rs}\,\phi_s$ satisfy the Lie algebra. These anti-commutation relations follow from a Lagrangian $\tilde\phi\cdot\dot\phi$, which leads to a propagator containing the step-function $\theta(t_1-t_2)$ (plus other terms depending on the boundary conditions) which is just what is needed to build the path-ordering in (\ref{3})\cite{Samuel:1978iy}. These two connections between fermionic variables and Lie algebras suggest we consider, instead of the functional integral in (\ref{2}), the simpler
\bq
&
-\int_0^\infty dT
\int
{\mathscr{D}}(w,\psi,\tilde\varphi,\varphi)\,{\cal N}e^{-\frac{1}{2}\int_0^{2\pi} dt \left(\frac{\dot w^2}{T}+\psi\cdot \dot\psi+\tilde\varphi\cdot\dot\varphi
+\tilde\varphi{\cal A}\varphi
\right)}\nonumber\\
&
\times\frac{1}{T}\int_0^{2\pi} dt\,\,\tilde\varphi\,
\psi\cdot\dot w\,
\psi\cdot\delta A
\,\varphi\,,
\label{4}\eq
The path-ordered exponential in (\ref{2}) can be picked out by a particular choice of boundary conditions an operator insertion and a choice of normalisation. We shall not follow this path, but rather we will find it more useful to consider the integral (\ref{4}) as it stands. The Lagrangian in this model has been analysed extensively \cite{Bala}-\cite{Salomonson:1977as}, mainly in canonical quantisation. It also arises naturally in models of tensionless strings interacting on contact: in \cite{James1}
and \cite{James2}, building on \cite{Mansfield:2011eq}, it was shown that the expectation value of the super-Wilson loop associated with a closed curve for an Abelian gauge theory is generated by the spinning string action integrated over world-sheets spanning the curve. Generalising this to the non-Abelian theory requires a way of extending the path-ordering associated with the boundary into the spanning world-sheet \cite{Next}, this can be achieved by the introduction of additional world-sheet fields with boundary values $\tilde\varphi$, $\varphi$. A full discussion of this string model is unnecessary here, save for the requirement that the Lie algebra generators be traceless. We re-write (\ref{4}) as
\be
\int_0^\infty dT
\int
{\mathscr{D}}(w,\psi)\,{\cal N}\,e^{-\frac{1}{2}\int_0^{2\pi} dt \left(\frac{\dot w^2}{ T}+\psi\cdot \dot\psi
\right)}\,\Omega\, Z[{\cal A}], \nonumber
\ee
where
\bq
&
\quad Z[{\cal A}]=\int
{\mathscr{D}}(\tilde\varphi,\varphi)\,e^{-\frac{1}{2}\int_0^{2\pi} dt \,\tilde\varphi\cdot{\cal D}\,\varphi
}\,,\quad
{\cal D}=\frac{d}{dt}+{\cal A}\,,
\nonumber\\
&
{\rm and}\quad\Omega=\frac{1}{T}\int_0^{2\pi} dt\,\psi\cdot\dot w\,\psi\cdot\delta A\,\frac{\delta}{\delta{\cal A}}\,.
\label{8}
\eq
To apply this to the Standard Model we use five pairs of $\tilde\varphi$ and $\varphi$, partitioned into sets of three and two to accommodate Lie algebra generators of $SU(3)\times SU(2)\times U(1)$ 
\be
(T^R)=i\left({\lambda^b}/2 , \quad {\sigma^a}/2, \quad
      1_2/2-1_3/3\right) \label{9}
\ee
where $\lambda$ and $\sigma$ are the $3\times 3$ Gell-Mann and $2\times 2$ Pauli matrices, and the form of the $U(1)$-generator is dictated by the tracelessness condition inherited from the underlying string model. Separate couplings are associated with each of the three algebras, but these have been implicitly absorbed into the gauge-field. Integrating over $\tilde\varphi$ and $\varphi$ gives $Z[{\cal A}]={\rm Det} \left(\,i{\cal D}\,\right)$
which we compute by solving the eigenvalue problem (with anti-periodic boundary conditions, like those on $\psi$)
\be
i{\cal D}\, v(t)=\mu\, v(t),\quad v(2\pi)=-v(0)
\ee
which has the solution $v(t)=g(t)\,v(0)\,\exp (-i\mu t)$ 
and $\mu=n+1/2+\log (\rho)/(2\pi i)$  
where $v(0)$ is required to be an eigenvector of the matrix $g(2\pi)$ with eigenvalue $\rho$, and $n$ is an integer. Now
\be
\prod_{n=-\infty}^\infty \left(1+\frac{\log \rho}{2\pi i(n+1/2)}\right)
=\frac{\sqrt\rho+1/\sqrt\rho}{ 2},
\ee
giving
\be
{\rm Det} \left(\,i{\cal D}\,\right)_{\rm ap}\propto
{\rm det}\left(\sqrt {g(2\pi)}+1/\sqrt {g(2\pi)}\right)\,.
\ee
${g(2\pi)}$ is block-diagonal with a $3\times 3$ piece
\be
\exp \left(\int_0^{2\pi}\frac{i}{ 3}{\cal A}^{U(1)}dt\right)\,
{\cal P}\,\exp \left(-\int_0^{2\pi}\frac{i}{ 2}{\cal A}^{SU(3)}\cdot\lambda dt\right)
\ee
which we denote by 
$e^{-i\theta/ 3}\,g_{SU(3)}$ and a  $2\times 2$ piece
\be
\exp \left(-\int_0^{2\pi}\frac{i}{ 2}{\cal A}^{U(1)}dt\right)\,
{\cal P}\,\exp \left(-\int_0^{2\pi}\frac{i}{ 2}{\cal A}^{SU(2)}\cdot\sigma dt\right)
\ee
which we denote by $e^{i\theta/2}\,g_{SU(2)}$
with $\theta=-\int_0^{2\pi} {\cal A}^{U(1)}dt$.
Furthermore the determinants can be expressed as sums of traces
so
\bq
&
{\rm Det} \left(\,i{\cal D}\,\right)_{\rm ap}\propto\left(e^{i\theta/2}+{\rm tr}\,g_{SU(2)}+e^{-i\theta/2}\right)\,\times
\nonumber\\
&
\left(e^{i\theta/2}+e^{i\theta/6}\,{\rm tr}\,g_{SU(3)}+e^{-i\theta/6}\,{\rm tr}\,g_{SU(3)}^\dagger+e^{-i\theta/2}\right)
\label{10}
\eq
The eigenvalues become $\mu=n+\log (\rho)/(2\pi i)$ if we impose periodic boundary conditions on the $\tilde\varphi$ and $\varphi$,  and
\be
\left(\prod_{n=-\infty}^{-1}\left(1+\frac{\log \rho}{2\pi in}\right)\right) 
{\log \rho}
\left(\prod_{n=1}^{\infty}\left(1+\frac{\log \rho}{2\pi in}\right)\right)={\sqrt\rho-1/\sqrt\rho},\nonumber
\ee
giving
\bq
&
{\rm Det} \left(\,i{\cal D}\,\right)_{\rm p}
\propto\left(e^{i\theta/2}-{\rm tr}\,g_{SU(2)}+e^{-i\theta/2}\right)\,\times
\nonumber\\
&
\left(-e^{i\theta/2}+e^{i\theta/6}\,{\rm tr}\,g_{SU(3)}-e^{-i\theta/6}\,{\rm tr}\,g_{SU(3)}^\dagger+e^{-i\theta/2}\right)
\label{11}
\eq
and if at the same time we impose periodic boundary conditions on $\psi$ the effect is to generate a $\gamma_5$ insertion.
Adding together the result of computing (\ref{8}) with anti-periodic boundary conditions on all fermions to that of imposing periodic boundary conditions results in the sum of 
(\ref{10}) and (\ref{11}) multiplied by $\gamma_5$ provided we {\it choose} the values of the normalisations of the functional integrals, ${\cal N}$, in the two cases appropriately. Multiplying out the terms results in the projection operators $P_L$ and $P_R$ multiplied by various exponentials of gauge fields:
\bq
&&
e^{i\theta}\,P_L+e^{i\theta/2}{\rm tr}\,g_{SU(2)}\,P_R+P_L
+e^{i2\theta /3}{\rm tr}\,g_{SU(3)}\,P_R
\nonumber\\
&&
+e^{i\theta /6}{\rm tr}\,g_{SU(3)}\,{\rm tr}\,g_{SU(2)}\,P_L
+e^{-i\theta/3}{\rm tr}\,g_{SU(3)}\,P_R\nonumber\\
&&
+e^{i\theta/3}{\rm tr}\,g_{SU(3)}^\dagger\,P_L
+
e^{-i\theta /6}{\rm tr}\,g_{SU(3)}^\dagger\,{\rm tr}\,g_{SU(2)}\,P_R\nonumber\\
&&
+e^{-i2\theta/3}{\rm tr}\,g_{SU(3)}^\dagger\,P_L+P_R
\nonumber\\&&
+e^{-i\theta/2}{\rm tr}\,g_{SU(2)}\,P_L
+e^{-i\theta}\,P_R
\eq
 From which we can read off the representations and chiralities of the fermions. Each term has a piece $e^{iY\theta}$ where $Y$ is the $U(1)$-charge. ${\rm tr}\,g_{SU(2)}$ represents the coupling of an $SU(2)$-doublet, ${\rm tr}\,g_{SU(3)}$ the coupling in the fundamental representation of $SU(3)$ and ${\rm tr}\,g_{SU(3)}^\dagger$ its complex conjugate. 
This set of twelve terms corresponds to the $U(1)$-charges, $SU(2)$, $SU(3)$ representations and chirality assignments of the fermions and their anti-particles in the Standard Model, augmented by a sterile neutrino (useful in modeling neutrino masses.):
$
\left(\bar l_R,\,\bar E_L,\,\bar\nu_R,\,U_R,\,Q_L,\,D_R,\,\bar D_R,\,\bar Q_L,\,\bar U_R,\,\nu_R,\,E_L,\,l_R
\right)
$, respectively.

\section{Concluding remarks}

We have described a simple generalisation of the world-line approach to chiral fermions that automatically produces the sum over the $SU(3)\times SU(2)\times U(1)$ representations and chiralities that occur in a single generation of the Standard Model augmented by a sterile right-handed neutrino. In the language of field theory the one-loop effective action in a background gauge-field is 
\bq
&{\rm log}\int{\mathscr{D}}(E_L,l_R,Q_L,U_R,D_R)
\exp \Big(-\int d^4x\,(E_L^\dagger\bar\sigma\cdot D\,E_L\nonumber\\
&+l_R^\dagger\sigma\cdot D\,l_R
+\nu_R^\dagger\sigma\cdot D\,\nu_R+Q_L^\dagger\bar\sigma\cdot D\,Q_L
\nonumber\\
&
+U_R^\dagger\sigma\cdot D\,U_R+D_R^\dagger\sigma\cdot D\,D_R)\Big)\,,\nonumber
\eq
where the gauge-covariant derivative, $D$, depends on the representation of the field it acts on.
By contrast the world-line expression for the variation of this effective action under a change of gauge-field is just
\be
\int {dT}\,{\mathscr{D}}(w,\psi,\tilde\varphi,\varphi)
\,
\Omega
\exp\left({-\frac{1}{2}\int dt\,\left(\frac{\dot w^2}{T}+\psi\cdot\dot\psi+\tilde\varphi\cdot{\cal D}\,\varphi\right)}\right)
\nonumber
\ee
summed over anti-periodic and periodic boundary conditions on the anti-commuting variables. ($\cal D$, $\cal A$ and $\Omega$ are defined in (\ref{8}) and (\ref{3}).) The information about representations and chiralities is generated by the functional determinant resulting from integrating out the additional anti-commuting world-line variables, $\tilde\varphi$ and $\varphi$. This model arises naturally in the context of tensionless strings with contact interactions where it becomes necessary to extend the notion of path-ordering along a closed curve into the body of a world-sheet spanning the curve.  

We made a number of choices. We chose the gauge group to be that of the Standard Model, with the generators of $SU(3)$ and $SU(2)$ as in (\ref{9}). The $U(1)$ generator was required to be traceless due to the underlying string model, but we chose its overall normalisation in (\ref{9}). We also added the result of applying periodic and anti-periodic boundary conditions on all the fermions. This is like a GSO projection, as might be anticipated given the connection with an underlying spinning string theory.

It is a pleasure to thank Anton Ilderton and James Edwards for discussions and STFC for support under the Consolidated Grants ST/J000426/1 and ST/L000407/1. This research is also supported by the Marie Curie network GATIS 
(gatis.desy.eu) of the European Union's Seventh Framework Programme FP7/2007-2013/ under REA Grant Agreement No 317089.





\end{document}